**To what extent is ChatGPT useful for language teacher lesson plan creation?**

Alex Dornburg*
University of North Carolina at Charlotte, USA (adornbur@charlotte.edu)

Kristin J. Davin
University of North Carolina at Charlotte, USA (kdavin@charlotte.edu)

**ORCID IDs**:

A. Dornburg:        0000-0003-0863-2283
K. Davin:           0000-0003-3590-7086

* Corresponding author:
Alex Dornburg
adornbur@charlotte.edu

**Abstract**

The advent of generative AI models holds tremendous potential for aiding teachers in the generation of pedagogical materials. However, numerous knowledge gaps concerning the behavior of these models obfuscate the generation of research-informed guidance for their effective usage. Here we assess trends in prompt specificity, variability, and weaknesses in foreign language teacher lesson plans generated by zero-shot prompting in ChatGPT. Iterating a series of prompts that increased in complexity, we found that output lesson plans were generally high quality, though additional context and specificity to a prompt did not guarantee a concomitant increase in quality. Additionally, we observed extreme cases of variability in outputs generated by the same prompt. In many cases, this variability reflected a conflict between 20th century versus 21st century pedagogical practices. These results suggest that the training of generative AI models on classic texts concerning pedagogical practices may represent a currently underexplored topic with the potential to bias generated content towards teaching practices that have been long refuted by research. Collectively, our results offer immediate translational implications for practicing and training foreign language teachers on the use of AI tools. More broadly, these findings reveal the existence of generative AI output trends that have implications for the generation of pedagogical materials across a diversity of content areas.

**Keywords:** ChatGPT; generative artificial intelligence; zero shot prompting; language teaching; historic bias; transformer model

1. **Introduction**

Language education is the only subject in which the language is both the content and the vehicle of instruction. Teachers typically face a range of proficiency levels in each class, with students developing their language skills at differing rates from their peers (Davin et al., 2014). To mitigate the challenges intrinsic to instruction in this content area, best practices for teaching languages have changed dramatically over the last century. For example, audiolingual methods in which learners memorized dialogues and grammar translation approaches in which students translated texts have been replaced with communicative approaches in which students use the language for real-world tasks (Lightbown & Spada, 2021). Such research-driven approaches have greatly improved the efficacy of classroom instruction. However, a consistent lack of resources continues to exacerbate teachers both in and out of the classroom. Foreign language education often exists as a marginalized content area with fewer resources than subjects such as math and science (Cawelti, 2006). In Anglophone contexts especially, programs typically have either outdated curricular materials and textbooks or none at all. In many schools, teachers are tasked with developing their own lesson plans and assessments (Schwartz, 2022). Thus, resources that can aid teachers in instructional design are critically needed.

The recent advent of readily available large language models (LLMs) such as ChatGPT are poised to dramatically impact how teachers approach instructional design. Such trends have already emerged in other fields such as consulting, where those trained in the use of AI tools have a documented 40% increase in work quality and 25% increase in task completion speed as compared to those who do not (Dell'Acqua et al., 2023). However, AI tools have not yet been widely embraced in foreign language programs. Many foreign language educators are cautious of this technology (Bekou et al., 2024; Kohnke et al., 2023), in part due to declining enrollments (Lusin et al., 2023) and recent headlines regarding the replacement of foreign language departments with commercial software that uses AI (e.g. Duolingo; Petit, 2023). This reticence and the nascent nature of generative AI is now creating a knowledge gap that threatens to further isolate foreign language teachers from this rapidly evolving field.

Understanding how to prompt AI models and the possible limitations of the resulting outputs are critical components of AI competency. Prompts are the interface between human intent and machine output, usually manifesting as questions or instructions given to an AI model with the goal of eliciting a specific response (Giray, 2023; Short & Short, 2023). In its simplest form, a user prompts an LLM like ChatGPT with a simple command and obtains an output (Wei et al., 2023). For LLMs to successfully complete complex tasks, the ability to engineer sophisticated prompts that contain a high level of specificity is critical (Giray, 2023; Zhong et al., 2023). Numerous AI prompt engineering strategies have been developed to guide generative AI solutions to desired outputs (Bozkurt & Sharma, 2023; White, Fu, et al., 2023). However, variability in outputs from users using the same prompt may be correct for one and contain errors for another. The degree to which such errors are to be expected or if they reflect general weaknesses of current generative AI models when generating pedagogical materials remains unknown.

The purpose of the present study was to provide the first assessment of how prompt specificity influences trends in output variability and weaknesses when using generative AI to write lesson plans for language teachers. We used zero-shot prompting, in which a user inputs a single prompt and does not engage in dialogue with the chatbot. This approach provides a baseline expectation of the model's output and is the most likely to be employed by non-AI specialists. We designed five prompts in which each subsequent prompt increased in the level of specificity. Additionally, we input each prompt 10 times to analyze the variability of outputs (i.e., lesson plans produced by ChatGPT) by scoring them against criteria based on best practices and the requirements of the most commonly required foreign language teacher licensure exam in the United States (i.e., edTPA). Using the resulting scores, we quantified the variation in outputs and assessed the overall strengths and weaknesses of ChatGPT for lesson plan creation. Collectively, these results provide essential guidance on the extent to which zero-shot prompting can be used for lesson plan creation.

## 2. Literature Review

### 2.1 What are Large Language Models?

A primary objective of natural language processing (NLP), a subfield of Artificial Intelligence, is to enable machines to understand, interpret, and generate human language for task performance (Chowdhary, 2020). The recent release of LLMs such as ChatGPT has placed unprecedented attention on our ability to create models that allow machines to mimic human language (Roe & Perkins, 2023). This ability is due in part to advances in deep learning, in which networks of nodes that mirror our conceptual understanding of human neural networks communicate and extract meaningful content from unstructured input data (Roumeliotis & Tselikas, 2023; Serban et al., 2016). Understanding of input text is made possible by pre-training models on aggregations of millions of pages of text from books, websites, articles, and other sources (Wu et al., 2023). This pre-training provides a foundational basis for capturing semantic nuances of human language that can be fine-tuned for a wide range of specific applications that span content creation (Cao et al., 2023), language translation (Gu, 2023; Li et al., 2023), and writing assistance (Bašić et al., 2023; Imran & Almusharraf, 2023) to name but a few. Central to the efficacy of such applications is the prompt of the user, which embeds task descriptions as input that guides the computational response of the AI model (Lee et al., 2023; White, Hays, et al., 2023).

### 2.2. Prompting in LLMs

Prompts act as the primary user-based input that LLMs such as ChatGPT respond to when generating output. A prompt may simply state a question or task command such as "Write a haiku about sharks." In response, ChatGPT will generate the haiku. If a haiku about any aspect of sharks is the desired output, then the user will have achieved their goal. However, such general examples are rarely the desired output. Instead, users of LLMs in professional settings have highly specialized tasks for which they need specific output. This need for specificity has led researchers to urge users of generative AI to understand and master fundamental concepts of prompt engineering to effectively leverage LLMs (Giray, 2023; Hatakeyama-Sato et al., 2023; Heston & Khun, 2023). Effective prompts are often comprised of four components (Giray, 2023):

(1) **Instruction**: A detailed directive (task or instruction) that steers the model's actions towards the intended result.
(2) **Context**: Supplementary data or context that supply the model with foundational understanding, thereby enhancing its ability to produce accurate outputs.
(3) **Input data**: This serves as the foundation of the prompt and influences the model's perception of the task. This is the query or information we seek to have the model analyze and respond to.
(4) **Output indicator:** This sets the output type and format, defining whether

a brief answer, a detailed paragraph, or another particular format or combination of formats are desired.

Implementing these components into inputs can aid in more readily guiding LLMs to accurate target outcomes (Giray, 2023; Jacobsen & Weber, 2023; Meskó, 2023).

*2.3 Variation in Outputs*

The ability of AI to generate non-deterministic outputs from the same prompt has been lauded as a major achievement, but it also underscores a need for caution. This ability enables the LLM to weigh the importance of words in a sentence and generate outputs based on probability distributions (Lubiana et al., 2023). Central to this architecture is the temperature parameter which acts as a dial for the model's creative output. At low temperature values, words with higher probabilities are chosen and the model output becomes more deterministic (Lubiana et al., 2023). At high temperature values, the model explores a broader range of possible responses that include novel and less expected outputs (Davis et al., 2024). However, even at low temperature settings near or at zero, models like ChatGPT have been found to return non-deterministic and sometimes erroneous results by chance. Jalil et al (2023) recently found that even at a temperature setting of zero, ChatGPT provided non-deterministic answers to simple prompts related to software curriculum nearly 10% of the time with an error rate of over 5%. As the default temperature setting for the public release of ChatGPT likely to be used by educators is around 0.7, this finding suggests that assessments of this tool in education should include the potential for non-determinism. Unfortunately, how variability in output responses to the same prompt impact the design of language teacher instructional materials remains unexplored.

**2.4** *Approaches to Prompting for Lesson Planning*

The promise of using LLMs for foreign language teacher material development was recognized not long after the public release of ChatGPT (Hong, 2023; Kartal, 2023; Koraishi, 2023; Kostka & Toncelli, 2023). Since then, a range of prompting techniques such as zero-shot (Kojima et al., 2022; Sanh et al., 2021), few-shot (Brown et al., 2020; Kojima et al., 2022), chain-of-thought (Wang et al., 2022), tree-of-thought (Yao et al., 2023), and even autotune solutions (Khattab et al., 2023) have been developed. However, whether teachers require training in complex prompting strategies for routine tasks remains unclear. Corp and Revelle (2022) explored the ability of eight pre-service elementary school teachers to use zero-shot prompting with ChatGPT for lesson plan creation and found it to be feasible after a short tutorial. Zero-shot approaches have repeatedly been shown to produce quality outputs when prompted effectively (Ateia & Kruschwitz, 2023; D. Hu et al., 2024; Y. Hu et al., 2023), including in an evaluation of materials generated for physics classes that found no statistical difference in output between zero-shot prompting and other more complex approaches (Yeadon & Hardy, 2024). More recently, Karaman & Göksu (2024) evaluated the effectiveness of lesson plans generated by ChatGPT using zero-shot prompting on third graders' mathematical performance over five weeks, noting a boost in math scores for the ChatGPT group. Although results such as this underscore the potential benefits of incorporating AI-developed lesson plans into the educational repertoire, whether the alignment of those plans with teaching objectives and standards is reliable has been questioned (Koraishi, 2023; Lo, 2023) leaves this question unanswered.

**3. Methods**

The present study sought to analyze the extent to which zero shot prompting was useful for language teacher lesson plan creation. Specifically, the study was guided by the following research questions:

R1. To what degree does increasing the specificity of prompts impact the structure and content of AI-generated lesson plans?
R2. How does the specificity of a prompt influence the consistency of AI-generated responses?
R3. Does ChatGPT demonstrate any overall strengths or weaknesses in lesson plan design, regardless of prompt specificity?

### 3.1 Approach

We used ChatGPT 4.0 (OpenAI, 2024) to assess how increasing prompt specificity impacted the alignment of outputs with lesson plan criteria that are given to pre-service L2 teachers during their training at the University of North Carolina at Charlotte. We designed five increasingly specific zero-shot prompts following general guidelines of prompt design (Giray, 2023) and recorded the resulting outputs. Each prompt was input into 10 unique chats to additionally assess the resulting non-determinism of outputs between the same prompt. The resulting 50 outputs were scored for the presence/absence of criteria and subject to a range of statistical analyses that allowed for the visualization of trends, assessment of variability between prompt groups, and a series of analyses in dissimilarity. These analyses aimed to reveal whether specific features of the prompt design yielded outputs that were more or less aligned with target criteria for the lesson plan design. This approach allowed us to provide an assessment of zero-shot prompting for language teacher lesson plan design.

### 3.2 Prompt Design

The five prompts iteratively built specificity toward constructing a lesson plan that aligns with target criteria used in the foreign language teacher licensure program at the University of North Carolina at Charlotte. The lesson plan template and scoring rubric aligned to the requirements of Pearson's edTPA, a performance assessment that teachers in North Carolina and many other states must pass for teacher licensure. Prompts were designed to adhere to the guidance of prompt design that include instruction, context, input data, and an output indicator (Giray, 2023). They iteratively increased in complexity. Table 1 shows the initial prompt (P.1) and the phrase or direction that was added in each subsequent prompt.

**Table 1.** Prompts

| Level | Prompt |
| --- | --- |
| P.1 | Design a 45-minute lesson for a high school Level I Spanish classroom with 30 students at the novice level of proficiency. The lesson objectives must be: (1) I can describe a typical Costa Rican restaurant; (2) I can ask and answer questions about how much something costs, (3) I can identify the foods served at a restaurant by looking at the menu. Please output a lesson plan based on your design. |
| P.2 | P.1 + …novice level of proficiency based on the ACTFL Proficiency guidelines. |
| P.3 | P.2 + Please output a lesson plan based on your design using the provided format (See Supplement 1). |
| P.4 | P.3 + Please address as many of the ACTFL world-readiness standards as possible in the lesson. |
| P.5 | P.4 + The lesson should include the following components (See Supplement 2). |

As Table 1 displays, P.1 provided a general case prompt that included features common to lesson planning. P.2 added specificity concerning the definition of proficiency based on the ACTFL proficiency guidelines (ACTFL, 2012), which are guidelines that describe what language users can do with the language in real-world situations at different proficiency levels (e.g., novice, intermediate, advanced). Prompt P.3 increased in specificity by adding a lesson plan format (See Supplemental Material 1), uploaded as a pdf for ChatGPT to complete, that a pre-service teacher would utilize as part of their training. P.4 included all components of P.1, P.2, and P.3, but added the condition that the lesson plan should address multiple ACTFL world-readiness standards. Finally, P.5 included a checklist of criteria that individually listed each component that the lesson plan should include (See Supplement 2).

### *3.3 Dataset creation*

All prompts were used as input for ChatGPTv 4.0 (OpenAI, 2024) with the resulting lesson plan output saved to a text file. Each prompt (P.1-P.5) was input ten times to capture non-deterministic output arising from the temperature parameter. Each prompt and prompt iteration occurred on new chats to ensure no influence of the prior prompt on the output. All text files were labeled by prompt. For example, P1.1 represented the first time Prompt 1 was entered, P1.2 represented the second time Prompt 1 was entered, and P5.5 represented the fifth time Prompt 5 was entered. In sum, 50 lesson plans, 10 for each of the five prompts, were generated. We then scored each output for the presence (1) or absence (0) of components indicated in P.5, yielding a binary presence/absence matrix for data analyses.

### *3.4 Statistical Analyses*

All statistical analyses were conducted in R v4.2.1 "Bird Hippy" (R Development Core Team, 2021). To investigate how increasing the specificity of prompts impacted the structure and content of AI-generated lesson plans (**R1**), we first assessed general trends in the presence or absence of target lesson components. Trends in the total number of target components captured by each prompt across iterations were visualized and group means compared using ANOVA. This allowed us to test if the prompt types were significantly different in terms of generating outputs more aligned with the target criteria. Pairwise t-tests between groups were conducted using the correction method of Benjamini and Hochberg (Benjamini & Hochberg, 1995) to mitigate the potential for false discovery. These analyses were additionally repeated on the word counts between prompt categories to assess if specificity abridged the resulting output.

As an ANOVA only offers a perspective on group means, we additionally utilized several statistical approaches to analyze the dissimilarity between prompt outputs that can reveal separation of outputs that may be masked by comparisons of group means. As euclidean distance measures exhibit known pathologies when handling presence/absence matrices (Ricotta & Podani, 2017), we quantified dissimilarity using Jaccard distances, which are appropriate for binary data (Hao et al., 2019), using the vegan v2.6.4 package (Dixon, 2003). To visualize overlap between prompt outputs, we used nonmetric multidimensional scaling (NMDS), treating prompts (**P.1-P.5**) as groups. This commonly used approach to visualizing overlap of clusters (Alberti & Wang, 2022; Dornburg et al., 2016; Saeed et al., 2019) relaxes assumptions of linearity in alternate ordination approaches such as principal components analyses. To ensure that NMDS ordination is a viable indicator of dissimilarity, we quantified stress values, ensuring stress values below 0.1 (Clarke, 1993). To assess if prompts formed statistically significant clusters as would be expected if prompt specificity greatly impacted output (**R1**), we used the adonis2 function in vegan to conduct a permutational multivariate analysis of variance (PERMANOVA) with 999 permutations (Anderson, 2001; McArdle & Anderson, 2001). This test allowed us to assess if the composition of the prompts differed in multivariate space. More specifically, a significant difference would suggest that the grouping by prompts explains variance in goal achievement, indicating that different prompts lead to statistically different patterns of goal achievement. To gain additional insight into the degree of separation between groups, we complemented the PERMANOVA test with an analysis of similarities (ANOSIM), which tests the null hypothesis that there are no differences between groups through a comparison of within and between group dissimilarities (Chapman & Underwood, 1999; Clarke, 1993). Using 999 permutations, mean ranks ($R$) were quantified, with values near 0 indicating high similarity and values near 1 indicating high dissimilarity (Chapman & Underwood, 1999; Lamb et al., 2021).

To quantify how the specificity of a prompt influences the consistency of AI-generated responses (**R2**), distances were visualized in R to look for patterns of distance increasing as a function of prompt specificity. We next employed hierarchical clustering (Singh et al., 2011; Vijaya et al., 2019) with the single, complete, and average linkage algorithms in vegan to perform hierarchical clustering on the dissimilarity indices. This allowed us to visualize the distance between prompts and the degree to which output from the same prompt was clustered (**R2**) (Vijaya et al., 2019). Under a scenario in which prompts continually returned highly similar output, we would expect a high degree of clustering between replicates. In contrast, if the output is highly variable, we would expect a high degree of convergence in output scoring between outputs generated by different prompts.

Finally, to assess if ChatGPT demonstrated any overall strengths or weaknesses in lesson plan design, regardless of prompt specificity (**R3**), we quantified which components were absent across each prompt (e.g., Cultural Connections; meaningful context; etc).

## 4. Results

### *4.1 To what degree does increasing the specificity of prompts impact the structure and content of AI-generated lesson plans?*

There was a marginally significant difference in overall trends of prompt scores between prompt groups ($p$=0.0442; F=2.668; DF=4), though significance was not supported in multiple test corrected pairwise tests ($p > 0.12$). Adding detailed instructions had the effect of reducing the spread of variance and in some cases raising the mean value of scores (**Figure 1**). However, the mean scores did not increase linearly as a function of prompt detail. When the lesson template was first provided in P.3, the mean score dropped from 21.2 out of 25 from P.2 to 19.8. For P.4, when the prompt included the directive to address the ACTFL world-readiness standards, the resulting score decreased from 19.8 (P.3) to 19.6 (P. 4), remaining lower than the average scores from P.2. However, the addition of the checklist of criteria in P.5 raised the mean to the overall highest of 21.6. In addition, adding specificity has a significant impact on the overall word count (*p*=0.0047; F=4.338; DF=4), with P.3 being significantly shorter from P.1 & P.2 (*p*=0.007 and *p*=0.018, respectively) and P.4 being significantly shorter from P.1 (*p*=0.048).

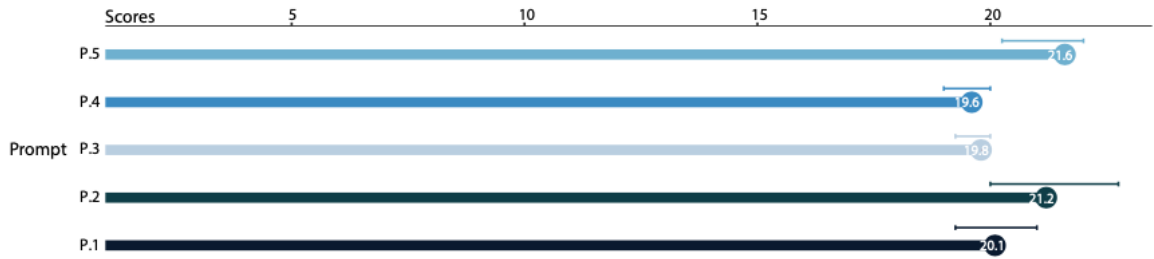

**Figure 1.** Average score of each prompt across iterations. Bars are shaded to correspond to specific prompts, bars above mean scores indicate 25% and 75% quantiles of the mean score.

NMDS-based visualizations of prompt output scores between prompting groups revealed several general aspects of the output alignments to scores (**Figure 2**). All prompt outputs shared some degree of overlap. However, there was also separation between substantial regions of the individual prompt clusters. This was particularly evident when comparing P.1+P.2 with the remaining prompts whose centroids largely occupied an alternate region of the NMDS space. This separation was supported by a PERMANOVA (F=3.1556; *p*=0.001) indicating significant differences in their multivariate profiles that supported that the prompt groups had significant effects on the outputs produced. The results of an ANOSIM complement the result of the PERMANOVA, again supporting significant differences between groups (R=0.141, *p*=0.001).

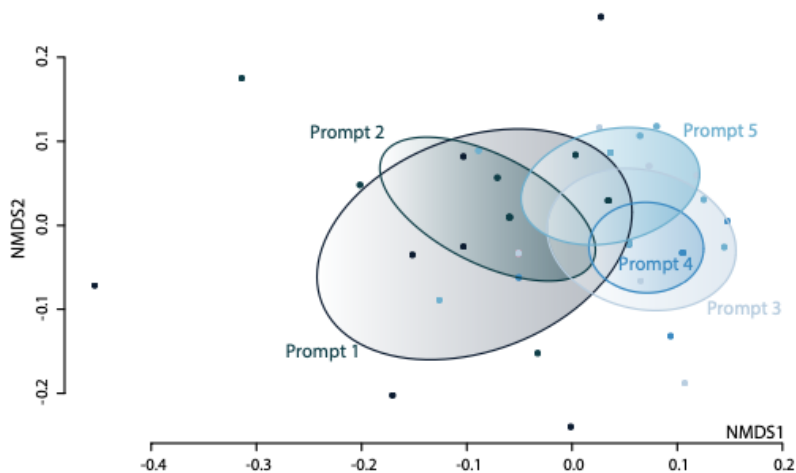

**Figure 2.** Visualization of the non-metric multidimensional scaling (NMDS) ordination plot of scores from all prompt groups showing group cluster overlap and separation. Ellipses represent estimated clusters and are shaded to correspond to the prompt group. Circles represent replicate scores and are shaded to the corresponding prompt group.

Qualitatively, the two prompt characteristics that most profoundly changed the ChatGPT output were the introduction of the lesson plan template in P.3 and the introduction of the scoring criteria in P.5. Once the lesson plan template was introduced, scores on the warm-up and teacher input portions of the lesson decreased dramatically. The average score on the warm-up, termed the "hook" in the template, revealed that P.1 (2.2 out of 3 possible points) and P.2 (2.5/3) scored much higher on the three portions of the checklist that corresponded to the rubric (See 2a-c in P.5) than P.3 (1.4), P.4 (1.0), and P.5 (1.4). ChatGPT repeatedly provided a warm-up that was some version of the following: "Quick video showcasing a vibrant Costa Rican restaurant scene, highlighting local foods, prices, and restaurant ambiance." Because that directive did not 1a) address the ACTFL standards or 2c) activate prior knowledge, it consistently scored lower.

The same was true for the Teacher Input. The average score for the four components related to the teacher input revealed that P.1 and P.5 both received perfect scores of 4 out of 4, unlike the prompts in between that scored 2.8/4 (P.2), 2/4 (P.3), and 2/4 (P.4). Lack of interaction and student engagement caused these low scores, even though the prompt for Teacher Input on the template stated, "Tip: This is where you introduce the new learning that addresses the Can-Do statement. You should engage students in interaction." Despite that directive, the lesson plans produced for P.3 and P.4 had teacher inputs with activities like, "Introduce vocabulary related to restaurants, foods commonly found in Costa Rica (e.g., "casado", "gallo pinto"), and phrases for asking about prices. Use images and realia to support learning".

*4.2 How does the specificity of a prompt influence the consistency of AI-generated responses?*

Scoring of prompt output revealed high variance in scores within prompt groups (Figure 3). For example, scores resulting from P.1 ranged from 23/25 to 16/25, with an average score of 20.1/25. Likewise, scores for P.5, which contained the scoring criteria ranged from a perfect score of 25 out of 25 to 20/25. In general, outputs generated from identical prompts varied by at least five elements. Corresponding to the high variance observed in the raw score, estimation of Jaccard distances provided little indication of distinct clustering by prompt type that would indicate strong dissimilarity between groups (**Figure 4**). Instead, within each prompt group, there were examples of highly divergent replicates as well as replicates that were highly similar to replicates from other prompt groups (**Figure 4A**). In other words, the convergence and divergence patterns observed in the distance matrix reveal a mixture of similarity and variability between and within prompt groups. For example, the tenth replicate of prompt 5 (5.10) was highly similar to the sixth replicate of prompt 2 (2.6), indicating the potential of a less detailed prompt output to converge in score with one generated using a higher detailed prompt by chance. In contrast, the third output generated using P.1 in independent chats (P1.3) was highly dissimilar to almost all other outputs (**Figure 4A**). This reveals the potential for a lack of rigid uniformity in how each distinct prompt type influenced outputs.

The dendrogram estimated using hierarchical clustering revealed a similar pattern to the raw distance matrix. Some prompt replicate outputs were highly dissimilar to the outputs from the other prompts (**Figure 4B**). Overall, there was some degree of differentiation between the prompt groups, with P.1 and P.2 having a higher distance on average from the P.3-P.5 replicates. However, the dendrogram also revealed numerous cases of convergence between prompt group replicates (Figure 4B). For example, 3 replicates from P.4 (P.4.9, P.4.5, P.4.3) were identical in scoring to a replicate from P.5 (P.5.3), two replicates from P.3 (P.3.4, P.3.2) and one replicate from P.1 (P.1.9). Similar cases of convergence in scoring groups were found throughout the dendrogram. Collectively, these results do not support the hypothesis that more specific prompts always lead to predictable and deterministic outputs, highlighting the importance of considering the variability inherent to AI-generated content when assessing prompt/output relationships.

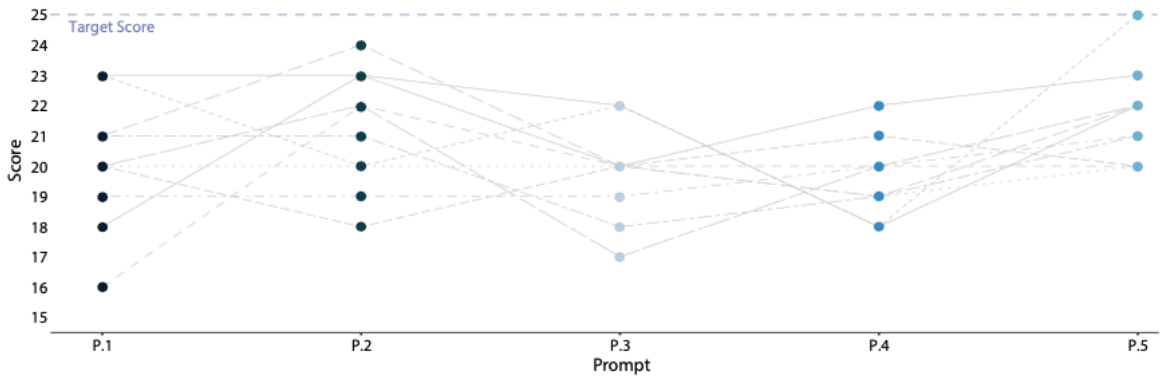

**Figure 3.** Alignment of prompt responses to rubric criteria. The resulting score of each prompt's (P.1-P.5) output relative to a perfect target score containing all 25 criteria. Circles indicate scores with dotted lines corresponding to each of the 10 prompt iterations. Circle shadings correspond to specific prompts.

12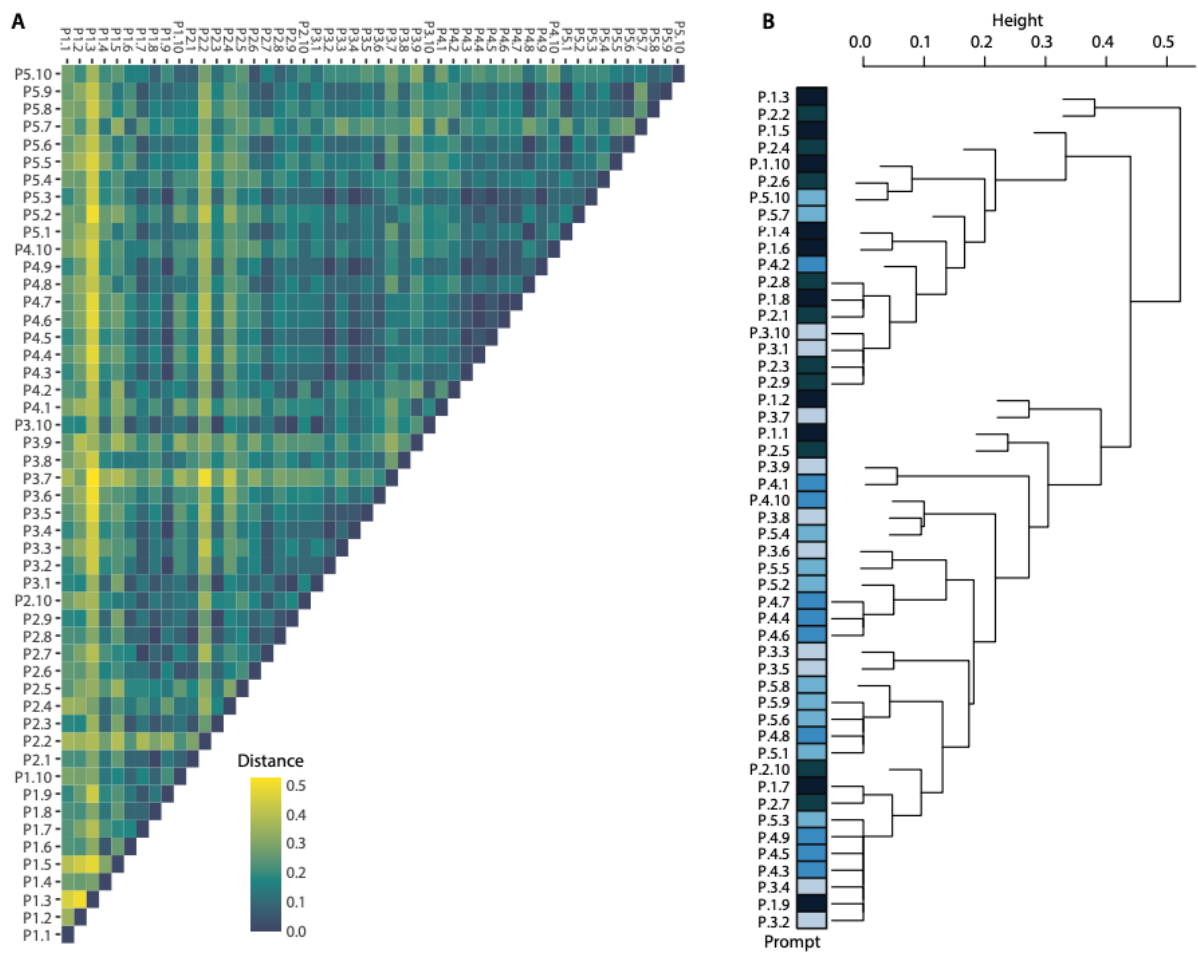

**Figure 4.** Distances between prompt output scores. (A) Heatmap depicting the computed distances between prompt replicate scores. Brighter shading indicates increased dissimilarity. **(B)** Dendrograms estimated using hierarchical clustering based on computed Jaccard distances. Shading in the centers corresponds to prompt (P.1-P.5). Dendrogram node heights correspond to the respective height. Prompt IDs in both panels indicate prompt group and replicate number (i.e., P.4.4= P.4, fourth replicate).

### *4.3 Does ChatGPT demonstrate any overall strengths or weaknesses in lesson plan design, regardless of prompt specificity?*

Assessing patterns of missing components in the scoring rubrics from prompt outputs revealed high heterogeneity between categories (**Figure 5**). Categories present in all outputs included meaningful context, teacher input aligning with lesson objectives, and activities appropriate to students' proficiency level. Several categories were also present in almost all cases except a few replicates of P.1 or P.2 including fostering student engagement, showing a connection to the learner's world, establishing a purpose for the lesson in the warm-up, and the integration of ACTFL standards into the closure and independent practice (**Figure 5**).

However, several categories were conspicuously absent between prompt groups. For example, cultural connection and ACTFL standards being integrated into teacher input were largely restricted to the outputs generated by P.5. Warm-up serving to activate prior knowledge and integration of the ACTFL standards into the focus and review were largely restricted to groups P.1 and P.2. Teacher input engaging learners in interaction was restricted largely to P.1, P.2, and P.5. In all of these examples of heterogeneity between prompt outputs, the appearance of rubric elements was often restricted to only around 50% of the outputs, indicating non-determinism in generated responses.

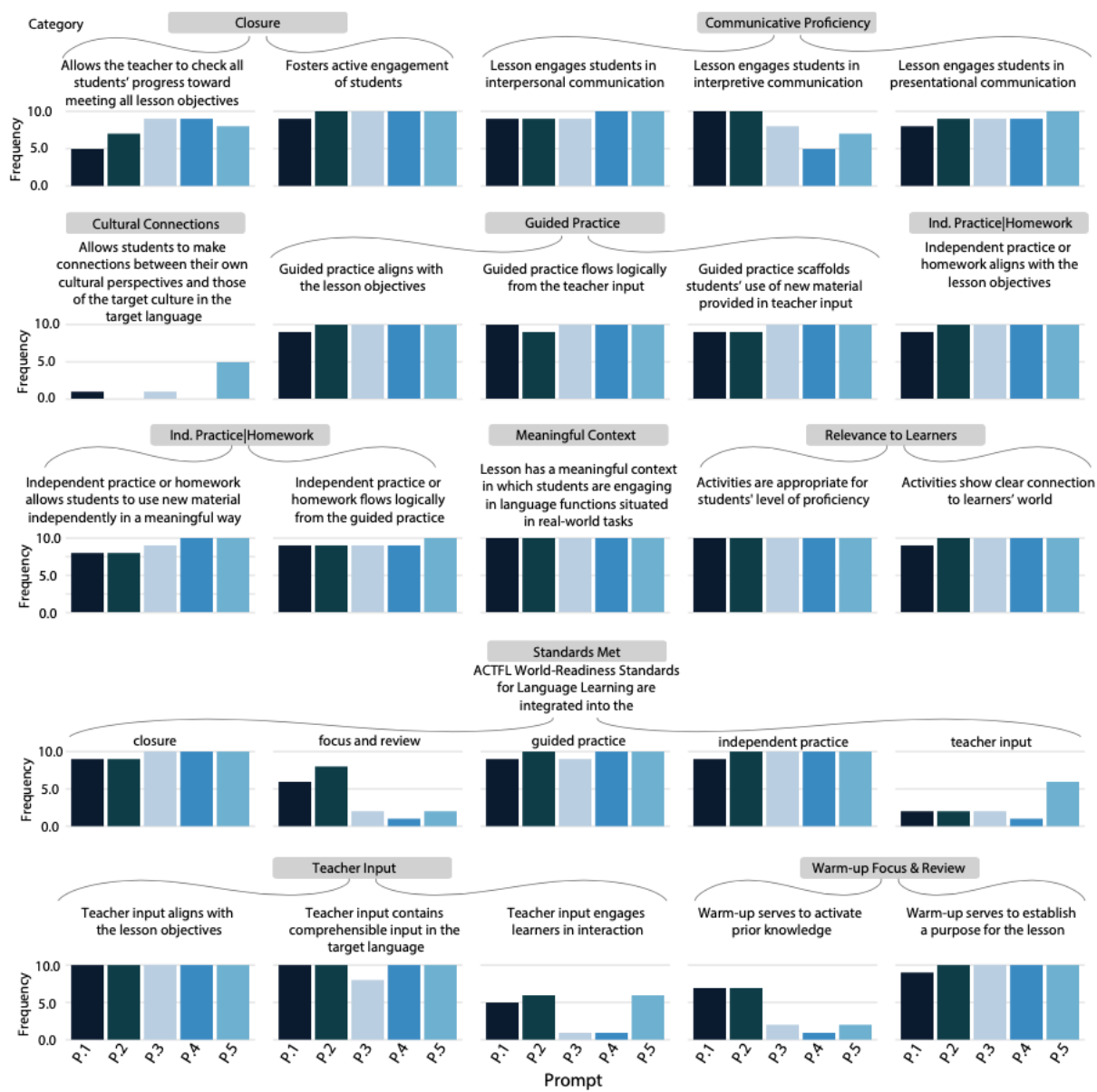

**Figure 5.** Summary of output scores by prompt group. 25 categories were scored as present or absent for each prompt replicate. Bar charts indicate the frequency of replicates meeting

scoring criteria. Bars are shaded to correspond to each prompt group.

## 5. Discussion

Our results demonstrate several significant aspects of ChatGPT's output in regards to prompt specificity, variability, and possible weaknesses that can guide effective usage. Overall, these results suggest a high degree of utility for developing foreign language teacher lesson plans, confirming existing research that zero-shot approaches can produce quality outputs (Ateia and Kruschwitz 2023; Hu et al. 2023, 2024). However, our results underscored that simply providing additional context and specificity to a prompt did not guarantee a concomitant increase in output quality. On the one hand, we observed a moderate degree of convergence in the outputs between the prompt categories, with several features of the scoring criteria present in all prompt outputs. On the other hand, we observed extreme cases of variability in which the same prompt yielded outputs perfectly or almost perfectly aligned with desired outcomes as well as prompts that missed numerous criteria. In several cases, this variability reflected outputs aligned with anachronistic pedagogical practices that no longer reflect best-practices. This suggests the presence of possible biases in the neural network stemming from training on historic data that may steer users towards research-rebuked teaching practices. Whether such biases permeate other aspects of instructional design requires additional study.

### *5.1 Increasing specificity does not always lead to increasing quality*

Determining the necessary specificity for a desired output is considered a critical aspect of prompting for AI models (Krause, 2023). However, we found that the relationship between the specificity of prompt and output score or generated lesson plans was not linear. For example, the average score of the output provided as a result of P.3 decreased from P.2 and remained virtually unchanged (0.2 difference) for P.4. In P.3, we provided a lesson plan template for the first time. This embedded description of the task guided the computational response of the AI model towards less detailed plans. Additionally, the opening section of a lesson plan, called the Warm-up or Focus and Review, was labeled as the hook (Wiggins & McTighe, 2005). ChatGPT seemed to interpret this terminology as input that is teacher-centered and does not require student interaction, which resulted in lower scores for related portions of the lesson plan. Consequently, once the lesson plan template was included as input beginning with P.3, the warmup/hook became almost formulaic. Though technically more specific, this decrease in score suggests that simply including a lesson plan template for context is not an effective strategy for increasing quality. Instead, including scoring criteria increased output quality Just as students are more effective when given scoring criteria along with the description of an assignment (Jonsson & Svingby, 2007), so too was ChatGPT. This was readily observed with the addition of the checklist with required criteria in P.5 that raised the mean to the overall highest of 21.6/25, suggesting that such input may be critical for optimizing effective AI-based lesson plan generation.

Translating these findings into practical implications for teachers highlights just how much context matters. When prompting ChatGPT to create foreign language lesson plans, teachers should include a meaningful context, lesson objectives, and the scoring criteria in the prompt. The inclusion of the first two in the prompt resulted in high scores for these categories across all prompts. The inclusion of scoring criteria in P.5 resulted in the highest average score on output for that prompt. If one's school or district requires the lesson in a particular format, then including that format can be useful as well. However, specialized terminology like the term

"hook" should be fully explained in the input. It is also important to note that a lesson plan template that works well for teachers' own instructional design may differ from one that works well for ChatGPT and may require iterative refinement. Teacher educators should experiment with using different templates for ChatGPT to determine which works best for their custom needs.

*5.2 Variable responses are an inherent feature of AI output*

Generative AI models break down prompts into textual representation and logic to parse the tasks required of them. As users add specificity and context to a prompt, this provides additional logic to guide the model towards desirable outputs (Giray, 2023). However, our work underscores that outputs from the same prompt are often not deterministic. In our case, scoring of prompt output revealed high variance in scores within prompt groups (Figure 1A). Simply by chance, some outputs from the same prompt missed over 25% of scoring criteria while others were over 90% complete. Similar variability and lack of deterministic output has been observed in other fields and is often attributed to similar weights between word pattern choices or the temperature value in the transformer model (Ouyang et al., 2023). Even at temperature settings of 0, models like ChatGPT have been found to return non-deterministic and erroneous outputs (Lubiana et al. 2023). Emphasizing to teachers the critical importance of evaluating and engaging with AI outputs should be an essential component of training AI competency, as instructional materials generated by teachers using the same prompt have the potential to widely vary in quality.

A possible approach to mitigating variability could involve training teachers to use additional follow-ups to improve output. In the present study, we did not engage in back-and-forth dialogue with the chatbot as this begins to introduce too much non-tractable variability. In contrast, once a chatbot produces a lesson plan, a teacher might follow-up with a prompt like, "Please engage students in interaction during the Teacher Input" or "Please script out questions for the teacher to ask during the Teacher Input" or "Please add a Closure in which the teacher asks students how dining practices might differ in Costa Rica and the United States." However, it is unlikely that such an approach would always yield desired outcomes. Based on the various temperature settings in generative models such as ChatGPT, there is no guarantee that outputs will be of a high quality in every instance. This is true even with refinement and iterative prompting and coincides with a general impediment to AI tool use: adopting AI content without critical evaluation (Dell'Acqua et al., 2023). For teachers, we advocate that this potential for a lack of engagement (Dell'Acqua et al., 2023) with AI tools and their output should be considered a key area of AI literacy and competency training. In practice, this would entail not only training teachers on how to write specific prompts, but also how to evaluate and revise the output produced. It is not clear how teachers currently using generative AI tools are evaluating outputs, and research on this topic is needed.

*5.3 Weaknesses in ChatGPT's output often reflected historic shifts in instructional design*

Generative AI models such as ChatGPT were trained on vast collections of text to develop mathematical associations between words that form the basis of each output. However, such training can induce algorithmic biases towards over-represented associations that do not reflect current knowledge or practices. Algorithmic bias may be particularly problematic for education, as there have been dramatic shifts in pedagogical practices between the 20th and 21st century. Our results reveal several potential sources of bias that can impact foreign language lesson plan generation. One of the most striking of those aspects involved including tasks that prompt students to make connections between their own cultural perspectives and those of the target culture in the target language. This checklist criteria were met in only five of the 50 lesson plans. Even when it was included in the input scoring rubric for P.5, output only included this objective 50% of the time. This output reflects the often-criticized historic practice in foreign language instruction of prioritizing the teaching of language at the expense of culture (Kramsch, 1991), and also the historic difficulties preservice teachers experienced with cultural comparisons as lesson components on their licensure exam (Hildebrandt & Swanson, 2014). This finding supports the view that the teaching of culture is an area within which current AI falls short (Kern, forthcoming), and suggests that this aspect of lesson planning requires human attention when using generative AI.

In addition to the integration of culture, theories of language acquisition and development have changed dramatically in the second half of this century (Lightbown & Spada, 2021). However, ChatGPT output related to engaging students in communicative activities in all sections of the lesson plan sometimes reflected approaches not aligned with contemporary research-backed practices. In particular, the audio-lingual method, a behaviorist approach popular in the 1970's in which learners listened to and repeated pre-scripted dialogues, seemed to influence some of the lesson plans. In one case, students were given a role play script to rehearse and present. In others, the plans asked the teacher to show flashcards for students to practice pronunciation. These historic practices appeared regardless of prompt specificity, even though the majority of output guided practice and independent practice activities were communicative and engaged students in interaction. Extrapolating from our results and considering the number of teachers who could use ChatGPT for lesson plan design, there is a high chance that similar relics of outdated methods will appear across outputs. Should users not look critically at generated output, this raises the concern that these atavisms might become pervasive features of pedagogical materials.

## 6. Conclusion

Many teachers and teacher educators are resistant to ChatGPT without realizing the intense pressure that language teachers face daily. Foreign language teachers often do not have access to high-quality curricula and there is a global trend towards abridging teacher preparation programs to eliminate barriers to the profession. Our results demonstrate that ChatGPT can streamline lesson planning thereby mitigating an aspect of professional burden. However, awareness of possible biases towards 20th century pedagogical practices that can occur by chance mark an urgent need in teacher AI literacy. Current rates of error that can be expected stochastically in foreign language instructional materials remain unknown because evaluations of models such as ChatGPT have been focused on single prompt outputs. Our work suggests that historical biases could be pervasive and should be expected to occur in output by chance. To avoid reintroducing research-rejected practices into modern curricula, it is essential that teachers be trained to modify and revise outputs.

**Ethical statement and competing interests (required)**
This work involved no human subjects. The authors declare no competing interests.

**Data availability statement**
Full prompts and ChatGPT generated outputs are available in the supplemental materials. Output scores and code used in the analyses are available on Zenodo (DOI: 10.5281/zenodo.11060097): https://zenodo.org/records/11060097.

**About the author(s)**
Author #1 is an Assistant Professor in the College of Computing and Informatics at the University of North Carolina at Charlotte. His research interests include AI applications, phylogenetics, evolutionary immunology, and comparative genomics.

Author #2 is a professor of foreign language education at the University of North Carolina at Charlotte. Her research interests include foreign language teacher development, language policies, and assessment.

**References**
ACTFL. (2012). *ACTFL proficiency guidelines.* https://www.actfl.org/publications/guidelines-and-manuals/actflproficiency-guidelines-2012
Alberti, M., & Wang, T. (2022). Detecting patterns of vertebrate biodiversity across the multidimensional urban landscape. *Ecology Letters*, *25*(4), 1027–1045. https://doi.org/10.1111/ele.13969
Anderson, M. J. (2001). A new method for non-parametric multivariate analysis of variance. *Austral Ecology*, *26*(1), 32–46. https://doi.org/10.1111/j.1442-9993.2001.01070.pp.x
Ateia, S., & Kruschwitz, U. (2023). *Is ChatGPT a biomedical expert? -- exploring the zero-shot performance of current GPT models in biomedical tasks*. https://doi.org/10.48550/ARXIV.2306.16108
Bašić, Ž., Banovac, A., Kružić, I., & Jerković, I. (2023). ChatGPT-3.5 as writing assistance in students' essays. *Humanities & Social Sciences Communications*, *10*(1). https://doi.org/10.1057/s41599-023-02269-7
Bekou, A., Ben Mhamed, M., & Assissou, K. (2024). Exploring opportunities and challenges of using ChatGPT in English language teaching (ELT) in Morocco. *Focus on ELT Journal*, *6*(1), 87–106. https://doi.org/10.14744/felt.6.1.7
Benjamini, Y., & Hochberg, Y. (1995). Controlling the false discovery rate: A practical and powerful approach to multiple testing. *Journal of the Royal Statistical Society*, *57*(1), 289–300. https://doi.org/10.1111/j.2517-6161.1995.tb02031.x
Bozkurt, A., & Sharma, R. C. (2023). Generative AI and prompt engineering: The art of whispering to let the genie out of the algorithmic world. *Asian Journal of Distance Education*, *18*(2), i – vii. https://asianjde.com/ojs/index.php/AsianJDE/article/view/749
Brown, T. B., Mann, B., Ryder, N., Subbiah, M., Kaplan, J., Dhariwal, P., Neelakantan, A., Shyam, P., Sastry, G., Askell, A., Agarwal, S., Herbert-Voss, A., Krueger, G., Henighan, T., Child, R., Ramesh, A., Ziegler, D. M., Wu, J., Winter, C., … Amodei, D. (2020). *Language models are few-shot learners*. https://doi.org/10.48550/ARXIV.2005.14165
Cao, Y., Li, S., Liu, Y., Yan, Z., Dai, Y., Yu, P. S., & Sun, L. (2023). *A comprehensive survey of*

*AI-generated content (AIGC): A history of generative AI from GAN to ChatGPT*. https://doi.org/10.48550/ARXIV.2303.04226

Cawelti, G. (2006). The side effects of NCLB. *Educational Leadership, 64*(3), 64.

Chapman, M. G., & Underwood, A. J. (1999). Ecological patterns in multivariate assemblages: Information and interpretation of negative values in ANOSIM tests. *Marine Ecology Progress Series*, *180*, 257–265. https://doi.org/10.3354/meps180257

Chowdhary, K. R. (2020). Natural language processing. In *Fundamentals of Artificial Intelligence* (pp. 603–649). Springer India. https://doi.org/10.1007/978-81-322-3972-7_19

Clarke, K. R. (1993). Non-parametric multivariate analyses of changes in community structure. *Australian Journal of Ecology*, *18*(1), 117–143. https://doi.org/10.1111/j.1442-9993.1993.tb00438.x

Corp, A, & Revelle, C. (2022). ChatGPT is here to stay: Using ChatGPT with student teachers for lesson planning. *The Texas Forum of Teacher Education*, *14*, 116–124.

Davin, K. J., Rempert, T. A., & Hammerand, A. A. (2014). Converting data to knowledge: One district's experience using large-scale proficiency assessment. *Foreign Language Annals*, *47*(2), 241–260. https://doi.org/10.1111/flan.12081

Davis, J., Van Bulck, L., Durieux, B. N., & Lindvall, C. (2024). The temperature feature of ChatGPT: Modifying creativity for clinical research. *JMIR Human Factors*, *11*, e53559. https://doi.org/10.2196/53559

Dell'Acqua, F., McFowland, E., Mollick, E. R., Lifshitz-Assaf, H., Kellogg, K., Rajendran, S., Krayer, L., Candelon, F., & Lakhani, K. R. (2023). *Navigating the jagged technological frontier: Field experimental evidence of the effects of AI on knowledge worker productivity and quality*. https://doi.org/10.2139/ssrn.4573321

Dixon, P. (2003). VEGAN, a package of R functions for community ecology. *Journal of Vegetation Science: Official Organ of the International Association for Vegetation Science*, *14*(6), 927–930. https://doi.org/10.1111/j.1654-1103.2003.tb02228.x

Dornburg, A., Lippi, C., Federman, S., Moore, J. A., Warren, D. L., Iglesias, T. L., Brandley, M. C., Watkins-Colwell, G. J., Lamb, A. D., & Jones, A. (2016). Disentangling the influence of urbanization and invasion on endemic geckos in tropical biodiversity hot spots: A case study of *Phyllodactylus martini* (Squamata: Phyllodactylidae) along an urban gradient in Curaçao. *Bulletin* , *57*(2), 147–164. https://doi.org/10.3374/014.057.0209

Giray, L. (2023). Prompt engineering with ChatGPT: A guide for academic writers. *Annals of Biomedical Engineering*, *51*(12), 2629–2633. https://doi.org/10.1007/s10439-023-03272-4

Gu, W. (2023). *Linguistically informed ChatGPT prompts to enhance Japanese-Chinese machine translation: A case study on attributive clauses*. https://doi.org/10.48550/ARXIV.2303.15587

Hao, M., Corral-Rivas, J. J., González-Elizondo, M. S., Ganeshaiah, K. N., Nava-Miranda, M. G., Zhang, C., Zhao, X., & von Gadow, K. (2019). Assessing biological dissimilarities between five forest communities. *Forest Ecosystems*, *6*(1). https://doi.org/10.1186/s40663-019-0188-9

Hatakeyama-Sato, K., Yamane, N., Igarashi, Y., Nabae, Y., & Hayakawa, T. (2023). Prompt engineering of GPT-4 for chemical research: what can/cannot be done? *Science and Technology of Advanced Materials: Methods*, *3*(1). https://doi.org/10.1080/27660400.2023.2260300

Heston, T., & Khun, C. (2023). Prompt engineering in medical education. *International Medical Education*, *2*(3), 198–205. https://doi.org/10.3390/ime2030019

Hildebrandt, S. A., & Swanson, P. (2014). World language teacher candidate performance on edTPA: An exploratory study. *Foreign Language Annals*, *47*(4), 576–591. https://doi.org/10.1111/flan.12117

Hong, W. C. H. (2023). The impact of ChatGPT on foreign language teaching and learning: opportunities in education and research. *Journal of Educational Technology and Innovation*, *5*(1). https://jeti.thewsu.org/index.php/cieti/article/view/103

Hu, D., Liu, B., Zhu, X., Lu, X., & Wu, N. (2024). Zero-shot information extraction from radiological reports using ChatGPT. *International Journal of Medical Informatics*, *183*, 105321. https://doi.org/10.1016/j.ijmedinf.2023.105321

Hu, Y., Chen, Q., Du, J., Peng, X., Keloth, V. K., Zuo, X., Zhou, Y., Li, Z., Jiang, X., Lu, Z., Roberts, K., & Xu, H. (2023). *Improving large language models for clinical named entity recognition via prompt engineering*. https://doi.org/10.48550/ARXIV.2303.16416

Imran, M., & Almusharraf, N. (2023). Analyzing the role of ChatGPT as a writing assistant at higher education level: A systematic review of the literature. *Contemporary Educational Technology*, *15*(4), ep464. https://doi.org/10.30935/cedtech/13605

Jacobsen, L. J., & Weber, K. E. (2023). *The promises and pitfalls of ChatGPT as a feedback provider in higher education: An exploratory study of prompt engineering and the quality of AI-driven feedback*. https://doi.org/10.31219/osf.io/cr257

Jalil, S., Rafi, S., LaToza, T. D., Moran, K., & Lam, W. (2023, April). ChatGPT and software testing education: Promises & perils. *2023 IEEE International Conference on Software Testing, Verification and Validation Workshops (ICSTW)*. 2023 IEEE International Conference on Software Testing, Verification and Validation Workshops (ICSTW), Dublin, Ireland. https://doi.org/10.1109/icstw58534.2023.00078

Jonsson, A., & Svingby, G. (2007). The use of scoring rubrics: Reliability, validity and educational consequences. *Educational research review, 2*(2), 130-144. https://doi.org/10.1016/j.edurev.2007.05.002

Karaman, M. R., & Göksu, İ. (2024). Are lesson plans created by ChatGPT more effective? An experimental study. *International Journal of Technology in Education*, *7*(1), 107–127. https://doi.org/10.46328/ijte.607

Kartal, G. (2023). Contemporary Language Teaching and Learning with ChatGPT. *Contemporary Research in Language and Linguistics (ISSN: 2980-2253)*, *1*(1). https://crlljournal.org/index.php/crll/article/view/10

Kern, R. (forthcoming). Twenty-first century technologies and language education: Charting a path forward. *Modern Language Journal, 108*(2).

Khattab, O., Singhvi, A., Maheshwari, P., Zhang, Z., Santhanam, K., Vardhamanan, S., Haq, S., Sharma, A., Joshi, T. T., Moazam, H., Miller, H., Zaharia, M., & Potts, C. (2023). *DSPy: Compiling declarative language model calls into self-improving pipelines*. https://doi.org/10.48550/ARXIV.2310.03714

Kohnke, L., Moorhouse, B. L., & Zou, D. (2023). ChatGPT for language teaching and learning. *RELC Journal*, *54*(2), 537–550. https://doi.org/10.1177/00336882231162868

Kojima, T., Gu, S. S., Reid, M., Matsuo, Y., & Iwasawa, Y. (2022). *Large language models are zero-shot reasoners*. https://doi.org/10.48550/ARXIV.2205.11916

Koraishi, O. (2023). Teaching English in the age of AI: Embracing ChatGPT to optimize EFL materials and assessment. *Language Education and Technology*, *3*(1). https://langedutech.com/letjournal/index.php/let/article/view/48

Kostka, I., & Toncelli, R. (2023). Exploring applications of ChatGPT to English language

teaching: Opportunities, challenges, and recommendations. *TESL-EJ*, *27*(3). http://files.eric.ed.gov/fulltext/EJ1409872.pdf

Kramsch, C. (1991). Culture in language learning: A view from the United States. In K. de Bot, R. B. Ginsberg, & C. Kramsch (Eds.), *Foreign language research in cross-cultural perspective* (pp. 217-240). John Benjamins Publishing Company.

Krause, D. (2023). Proper generative AI prompting for financial analysis. *SSRN Electronic Journal*. https://doi.org/10.2139/ssrn.4453664

Lamb, A. D., Lippi, C. A., Watkins-Colwell, G. J., Jones, A., Warren, D. L., Iglesias, T. L., Brandley, M. C., & Dornburg, A. (2021). Comparing the dietary niche overlap and ecomorphological differences between invasive geckos and a native gecko competitor. *Ecology and Evolution*, *11*(24), 18719–18732. https://doi.org/10.1002/ece3.8401

Lee, U., Jung, H., Jeon, Y., Sohn, Y., Hwang, W., Moon, J., & Kim, H. (2023). Few-shot is enough: Exploring ChatGPT prompt engineering method for automatic question generation in English education. *Education and Information Technologies*. https://doi.org/10.1007/s10639-023-12249-8

Li, B., Kou, X., & Bonk, C. J. (2023). Embracing the disrupted language teaching and learning field: Analyzing YouTube content creation related to ChatGPT. *Languages*, *8*(3), 197. https://doi.org/10.3390/languages8030197

Lightbown, P. M., & Spada, N. (2021). *How languages are learned* (5th Ed). Oxford University Press.

Lo, C. K. (2023). What is the impact of ChatGPT on education? A rapid review of the literature. *Education in Science: The Bulletin of the Association for Science Education*, *13*(4), 410. https://doi.org/10.3390/educsci13040410

Lubiana, T., Lopes, R., Medeiros, P., Silva, J. C., Goncalves, A. N. A., Maracaja-Coutinho, V., & Nakaya, H. I. (2023). Ten quick tips for harnessing the power of ChatGPT in computational biology. *PLoS Computational Biology*, *19*(8), e1011319. https://doi.org/10.1371/journal.pcbi.1011319

Lusin, N., Peterson, T., Sulewski, C., & Zafer, R. (2023). Enrollments in languages other than English in US institutions of higher education, fall 2021. Modern Language Association of America.

McArdle, B. H., & Anderson, M. J. (2001). Fitting multivariate models to community data: A comment on distance-based redundancy analysis. *Ecology*, *82*(1), 290–297. https://doi.org/10.1890/0012-9658(2001)082[0290:FMMTCD]2.0.CO;2

Meskó, B. (2023). Prompt engineering as an important emerging skill for medical professionals: Tutorial. *Journal of Medical Internet Research*, *25*, e50638. https://doi.org/10.2196/50638

OpenAI. (2024). ChatGPT (Mar 20 version) [Large language model]. https://chat.openai.com/chat

Ouyang, S., Zhang, J. M., Harman, M., & Wang, M. (2023). *LLM is like a box of chocolates: The non-determinism of ChatGPT in code generation*. https://doi.org/10.48550/ARXIV.2308.02828

Pettit, E. (2023). Scholars see dangerous precedent in West Virginia U.'s plan to cut foreign languages. The Chronicle of Higher Education. 18th August 2023. https://www.chronicle.com/article/scholars-see-dangerous-precedent-in-west-virginia-u-s-plan-to-cut-foreign-languages

R Development Core Team. (2021). *R: A language and environment for statistical computing*. R Foundation for Statistical Computing, Vienna, Austria. https://www.R-project.org/

Ricotta, C., & Podani, J. (2017). On some properties of the Bray-Curtis dissimilarity and their ecological meaning. *Ecological Complexity*, *31*, 201–205. https://doi.org/10.1016/j.ecocom.2017.07.003

Roe, J., & Perkins, M. (2023). "What they"re not telling you about ChatGPT': Exploring the discourse of AI in UK news media headlines. *Humanities & Social Sciences Communications*, *10*(1). https://doi.org/10.1057/s41599-023-02282-w

Roumeliotis, K. I., & Tselikas, N. D. (2023). ChatGPT and open-AI models: A preliminary review. *Future Internet*, *15*(6), 192. https://doi.org/10.3390/fi15060192

Saeed, N., Nam, H., Haq, M. I. U., & Muhammad Saqib, D. B. (2019). A survey on multidimensional scaling. *ACM Computing Surveys*, *51*(3), 1–25. https://doi.org/10.1145/3178155

Sanh, V., Webson, A., Raffel, C., Bach, S. H., Sutawika, L., Alyafeai, Z., Chaffin, A., Stiegler, A., Scao, T. L., Raja, A., Dey, M., Bari, M. S., Xu, C., Thakker, U., Sharma, S. S., Szczechla, E., Kim, T., Chhablani, G., Nayak, N., … Rush, A. M. (2021). *Multitask prompted training enables zero-shot task generalization*. https://doi.org/10.48550/ARXIV.2110.08207

Serban, I. V., Lowe, R., Charlin, L., & Pineau, J. (2016). *Generative deep neural networks for dialogue: A short review*. https://doi.org/10.48550/ARXIV.1611.06216

Short, C. E., & Short, J. C. (2023). The artificially intelligent entrepreneur: ChatGPT, prompt engineering, and entrepreneurial rhetoric creation. *Journal of Business Venturing Insights*, *19*(e00388), e00388. https://doi.org/10.1016/j.jbvi.2023.e00388

Singh, W., Hjorleifsson, E., & Stefansson, G. (2011). Robustness of fish assemblages derived from three hierarchical agglomerative clustering algorithms performed on Icelandic groundfish survey data. *ICES Journal of Marine Science: Journal Du Conseil*, *68*(1), 189–200. https://doi.org/10.1093/icesjms/fsq144

Vijaya, Sharma, S., & Batra, N. (2019, February). Comparative study of single linkage, complete linkage, and ward method of agglomerative clustering. *2019 International Conference on Machine Learning, Big Data, Cloud and Parallel Computing (COMITCon)*. 2019 International Conference on Machine Learning, Big Data, Cloud and Parallel Computing (COMITCon), Faridabad, India. https://doi.org/10.1109/comitcon.2019.8862232

Wang, X., Wei, J., Schuurmans, D., Le, Q., Chi, E., Narang, S., Chowdhery, A., & Zhou, D. (2022). *Self-consistency improves chain of thought reasoning in language models*. https://doi.org/10.48550/ARXIV.2203.11171

Wei, X., Cui, X., Cheng, N., Wang, X., Zhang, X., Huang, S., Xie, P., Xu, J., Chen, Y., Zhang, M., Jiang, Y., & Han, W. (2023). *Zero-shot information extraction via chatting with ChatGPT*. https://doi.org/10.48550/ARXIV.2302.10205

White, J., Fu, Q., Hays, S., Sandborn, M., Olea, C., Gilbert, H., Elnashar, A., Spencer-Smith, J., & Schmidt, D. C. (2023). *A prompt pattern catalog to enhance prompt engineering with ChatGPT*. https://doi.org/10.48550/ARXIV.2302.11382

White, J., Hays, S., Fu, Q., Spencer-Smith, J., & Schmidt, D. C. (2023). *ChatGPT prompt patterns for improving code quality, refactoring, requirements elicitation, and software design*. https://doi.org/10.48550/ARXIV.2303.07839

Wiggins, G., & McTighe, J. (2005). *Understanding by design.* ASCD.

Wu, T., He, S., Liu, J., Sun, S., Liu, K., Han, Q.-L., & Tang, Y. (2023). A brief overview of ChatGPT: The history, status quo and potential future development. *IEEE/CAA Journal of Automatica Sinica*, *10*(5), 1122–1136. https://doi.org/10.1109/jas.2023.123618

Yao, S., Yu, D., Zhao, J., Shafran, I., Griffiths, T. L., Cao, Y., & Narasimhan, K. (2023). *Tree of thoughts: Deliberate problem solving with large language models*. https://doi.org/10.48550/ARXIV.2305.10601

Yeadon, W., & Hardy, T. (2024). The impact of AI in physics education: A comprehensive review from GCSE to university levels. *Physics Education*, *59*(2), 025010. https://doi.org/10.1088/1361-6552/ad1fa2

Zhong, Q., Ding, L., Liu, J., Du, B., & Tao, D. (2023). *Can ChatGPT understand too? A comparative study on ChatGPT and fine-tuned BERT*. https://doi.org/10.48550/ARXIV.2302.10198

**Supplement 1**. Lesson plan output format uploaded as part of the input prompt

| Activity | Description of Activities and Setting | ACTFL standards addressed (1.1, 1.2, or 1.3) | Time |
|---|---|---|---|
| 1. Hook | | | |
| 2. Statement of Objective/Can-Do statement for Students<br>*Tip: "The student will be able to …" Rephrase the daily objective in "student language" if needed. This may just be repeated from above.* | | | |
| 3. Teacher Input<br>*Tip: This is where you introduce the new learning that addresses the Can-Do statement. You should engage students in interaction.* | | | |
| 4. Guided Practice<br>*Tip: This is where you guide students in scaffolded and supported practice of the new material you just introduced. This should involve 1 of the 3 modes of communication.* | | | |
| 5. Independent Practice | | | |
| *Tip: This is where students work in pairs, small groups, or individually to practice the new material provided in the input. This should involve 1 of the 3 modes of communication.* | | | |
| 6. Closure<br>*Tip: This should close the lesson and allow you to collect data on whether all students have met your lesson objective. Closure should involve at least 1 of the 3 modes of communication* | | | |

Materials/Technology:

**Supplement 2.** Scoring Criteria

    1) Standards met
        a) ACTFL World-Readiness Standards for Language Learning are integrated into the focus and review.
        b) ACTFL World-Readiness Standards for Language Learning are integrated into the teacher input.
        c) ACTFL World-Readiness Standards for Language Learning are integrated into the guided practice.
        d) ACTFL World-Readiness Standards for Language Learning are integrated into the independent practice.
        e) ACTFL World-Readiness Standards for Language Learning are integrated into the closure.

    2) Meaningful Context
        a) Lesson has a meaningful context in which students are engaging in language functions situated in real-world tasks.

    3) Warm-Up/Focus & Review
        a) Warm-up serves to activate prior knowledge.
        b) Warm-up serves to establish a purpose for the lesson.

    4) Teacher Input (I do with interaction)

    a) Teacher input aligns with the lesson objectives.
    b) Teacher input engages learners in interaction.
    c) Teacher input contains comprehensible input in the target language.

5) Guided Practice (We do)
    a) Guided practice aligns with the lesson objectives.
    b) Guided practice flows logically from the teacher input.
    c) Guided practice scaffolds students' use of new material provided in teacher input.

6) Independent Practice/Homework (You do)
    a) Teacher input aligns with the lesson objectives.
    b) Independent practice or homework flows logically from the guided practice.

7) Closure
    a) Closure allows the teacher to check all students' progress toward meeting all lesson objectives.
    b) Fosters active engagement of students.

8) Cultural Connections
    a) Lesson will allow students to make connections between their own cultural perspectives and those of the target culture in the target language.

9) Communicative Proficiency
    a) Lesson engages students in interpersonal communication.
    b) Lesson engages students in interpretive communication.
    c) Lesson engages students in presentational communication.

10) Relevance to Learners
    a) Activities show a clear connection to learners' world.
    b) Activities are appropriate for students' level of proficiency.